\documentstyle[aps,epsf,twocolumn]{revtex}

\newcommand{\be}{\begin{eqnarray}}
\newcommand{\ee}{\end{eqnarray}}

\begin{document}

\draft
\title{\bf  Is Quantization of QCD Unique at the Non-Perturbative Level ?}

\author{{\bf Hidenaga Yamagishi}$^1$  
and {\bf Ismail Zahed}$^2$}

\address{$^1$4 Chome 11-16-502, Shimomeguro, Meguro, Tokyo, Japan. 153;\\
$^2$Department of Physics, SUNY, Stony Brook, New York 11794, USA.}
\date{\today}
\maketitle

\begin{abstract}
We find that QCD in covariant gauge yields zero for the topological 
susceptibility, even at the nonperturbative level. The result is derived in two 
ways, one using translational invariance, and the other using the BRST 
Hamiltonian. Comparison with the canonical formalism suggests that QCD is not
uniquely defined at the nonperturbative level. Supporting evidence is also 
provided in 1+1 dimensions. Our results imply that the strong CP problem admits 
a trivial resolution in covariant gauge, but obstacles remain for the
U(1) problem.

\end{abstract}
 \narrowtext

{\bf 1. \,\,\,}
QCD is currently by far the favored theory of strong interactions. At the 
perturbative level, it fares well : the theory is well defined and agrees
with experiment. The situation is less satisfactory at the non-perturbative
level. Besides the well-known problems of demonstrating confinement and chiral 
symmetry breaking, there is a growing body of evidence that the theory is not 
uniquely defined at the non-perturbative level. First, dimensional transmutation
implies that the QCD perturbation series is not Borel summable, so the series 
does not uniquely define the theory. The Gribov ambiguity \cite{GRIBOV} opens 
the possibility that covariant and canonical quantization are inequivalent. 
This note adds another piece of evidence : BRST quantization implies that 
physics is automatically independent of the vacuum angle $\theta$, in contrast
with the canonical formalism. In particular, the topological susceptibility
$\chi$ is identically zero in covariant gauge, which has significant 
consequences for the U(1) problem. Evidence for a non-unique $\theta$ 
dependence is also found in 1+1 dimensions.

\vskip .5cm
{\bf 2.\,\,\,}
It is simple to see that $\chi$ is zero in covariant gauge \cite{YAM}.
As well known, the topological term in the QCD Lagrangian $\theta\,\,\Xi$
is a total divergence,
\be
\Xi=(g^2/32\pi^2)F^a_{\mu\nu}\tilde F^{\mu\nu a} =\partial^{\mu} K_{\mu}
\label{X1}
\ee
where
\be
K_{\mu} =\frac {g^2}{32\pi^2} \epsilon_{\mu\nu\rho \sigma} 
(A^{\nu a} F^{\rho \sigma a} -\frac g3 f^{abc} A^{\nu a} A^{\rho b} 
A^{\sigma c} )
\label{1}
\ee
is the Loos-Chern-Simons current \cite{LOOS}. In covariant gauge 
\be
<\theta | K_{\mu} (x) |\theta > = &&
<\theta | e^{iP\cdot x} \, K_{\mu} (0) \, e^{-iP\cdot x} |\theta >\nonumber\\
=&& <\theta | K_{\mu} (0) |\theta >
\label{X8}
\ee
is constant (just as for the Higgs field), 
where $|\theta >$ are the $\theta$ vacua and 
$P_{\mu}$ is the energy-momentum operator. Hence,
$<\theta |\Xi |\theta > = \partial^{\mu} <\theta | K_{\mu} |\theta > =0$
for any $\theta$, and $\chi =\partial <\theta|\Xi |\theta >/\partial\theta =0$
as well. 

It is important that the derivation fails to go through in the canonical 
formalism. The vacuum state then obeys Gauss' law, so the vacuum expectation 
value $<K_{\mu} >_{\theta}$ is ambiguous like $<x>_p$ in 
quantum mechanics. However, no such ambiguity exists in covariant gauge.
Since $\theta$ can be taken as a coupling constant, the so-called $\theta$ 
superselection rule trivially operates and a single Hilbert 
space suffices to describe physics for fixed $\theta$, including gauge-variant
operators. This space has indefinite metric, so one must further 
construct the physical Hilbert space out of it \cite{KUGO}. Nevertheless, 
as familiar since Gupta and Bleuler, the 
norm of the vacuum state in the original Hilbert space is equal to the norm in 
the physical Hilbert space, which is one. Hence $|\theta>$ is a proper element 
of the original Hilbert space, and $<\theta | K_{\mu} |\theta >$ is well 
defined. 

Our analysis applies to any formulation of QCD where the vacuum expectation 
value of a gauge-variant operator is well-defined and translation invariant.

\vskip .5cm
{\bf 3.\,\,\,}
The previous result may also be derived by another argument which is more 
formal but yields stronger results as well as some new insight.

In classical theory, adding a total diverence to the Lagrangian does not 
affect the physics, since the equations of motion are unchanged. However, 
this does not mean that the Hamiltonian remains the same; rather it undergoes 
a canonical transformation. 

Similarly, the BRST Hamiltonian for different 
vacuum angles is related by a unitary transformation
$H(\theta ) = e^{i\theta {\bf X}} H(0) e^{-i\theta {\bf X}}$ where
${\bf X} = \int d^3x K_0 $. This may be explicitly checked in covariant gauge, 
or in general derived from the quantum action principle,
\be
\delta_{\theta} <2|1> = i\delta\theta\,\,<2|\left( {\bf X} (2) -{\bf X} 
(1)\right) |1>
\label{qap}
\ee
where $1,2$ refer to spacelike hypersurfaces. In the former case, one must make 
sure to write the Hamiltonian in terms of canonical variables \cite{WITTEN}.

The question therefore is whether unitarily equivalent Hamiltonians can give 
inequivalent physics.

Let $|0 >$ be the physical ground state of 
$H( 0)$ in the BRST formalism. It must obey \cite{KUGO}
\be
H(0) |0> = 0 \qquad\qquad {\bf Q}_{\rm BRST} |0> =0
\label{2}
\ee
where ${\bf Q}_{\rm BRST}$ is the BRST charge.
Since ${\bf X}$ is invariant under infinitesimal gauge transformations, it is 
invariant under BRST transformations. Hence,
\be
H(\theta ) e^{i\theta {\bf X}} |0> = 0 \qquad\qquad 
{\bf Q}_{\rm BRST} e^{i\theta {\bf X}} |0> =0
\label{3}
\ee
which means that $e^{i\theta {\bf X}} |0>$ is the physical ground state of 
$H(\theta )$ with the same energy. The $\theta$-independence of the vacuum 
energy density ${\cal E}$ implies that 
$\chi=-\partial^2{\cal E}/\partial\theta^2 =0$ as before.

A further consequence is that strong CP violation is absent in the BRST 
formalism for any value of $\theta$, provided
it is absent for $\theta=0$.  If CP(0) commutes  with $H(0)$ and 
${\bf Q}_{\rm BRST}$ with $CP(0) |0>= |0>$, 
then CP($\theta$) $=e^{i\theta {\bf X}}$ CP(0) 
$e^{-i\theta {\bf X}}$ commutes with $H(\theta)$ and ${\bf Q}_{\rm BRST}$,
with CP($\theta$) $e^{i\theta {\bf X}} \, |0>= e^{i\theta {\bf X}} |0>$.
It is easy to extend these arguments to show that $H(\theta )$ is physically
equivalent to $H(0)$ in the BRST formalism. The physical states and observables 
are in one-to-one correspondence, and the matrix elements are the same.

The results may again be contrasted with the canonical formalism. The Hamiltonian 
then still has the structure $H(\theta ) =e^{i\theta {\bf X}} H(0) 
e^{-i\theta {\bf X}}$, but the subsidiary condition is altered to \cite{NOTE}
\be
{\bf U} [\Omega ] |{\rm phys}> = |{\rm phys}>
\label{X5}
\ee
where ${\bf U} [\Omega ]$ is the operator implementing the proper
gauge transformation $\Omega$.  If $\Omega$ is `large', ${\bf X}$ is shifted 
and $e^{i\theta {\bf X}}|0>$ is no longer a physical state unless $\theta$ is
an integer multiple of $2\pi$. Hence, Hamiltonians with different $\theta$
are generally expected to be inequivalent modulo $2\pi$. This is in agreement 
with the general wisdom, although we are unaware of any explicit calculations
for QCD in the canonical formalism \cite{VEGA}.

\vskip 0.5cm
{\bf 4.\,\,\,}
The Hamiltonian analysis above suggests that the source of the probable
inequivalence is that BRST invariance 
\be
{\bf Q}_{\rm BRST} |{\rm phys}>=0
\label{BB}
\ee
mimics Gauss' law, but not invariance under `large' gauge transformations. As a 
check, QED is independent of $\theta$, since there are no  `large' gauge 
transformations. Also in the Polyakov model \cite{POLYAKOV}, the analogue
of ${\bf X}$ is fully gauge invariant, so it commutes with either constraints 
(\ref{X5}) or (\ref{BB}). It follows that physics is again independent of 
$\theta$, in agreement with explicit computation \cite{VERGELES}.

Let us therefore entertain the possibility that the BRST formalism for QCD 
is missing some extra constraints. First, it will not do to simply impose 
(\ref{X5}) with `large' $\Omega$ in the BRST formalism. The BRST Hamiltonian is 
not invariant under ordinary gauge transformations, so the new constraint will 
not be preserved under time evolution. Furthermore, an extra constraint would
not help as far as covariant gauge is concerned. The
vacuum would remain normalizable and translational invariant, so $<\theta |\Xi 
|\theta>$ would remain zero by (\ref{X8}). 

A different approach is to replace (\ref{2}) by the weaker condition
\be
H(0) |0> = {\bf Q}_{\rm BRST} |\Lambda > \qquad\qquad {\bf Q}_{\rm BRST} |0> =0
\label{2X}
\ee
where $|\Lambda >$ is some vector. However, this upsets translational 
invariance for the vacuum expectation value of gauge-variant operators, so the
resulting formalism is no longer covariant gauge.

The above arguments do not completely rule the possibility that a suitable 
modification of the BRST formalism may bring it in line with the canonical 
formalism. However, the arguments are sufficient to show that our basic 
result will remain unaltered. Covariant gauge gives 
$<\theta | \Xi |\theta > =0$, even if the other (yet hypothetical) formalism 
gives $<\theta | \Xi |\theta > \neq 0$.

\vskip .5cm
{\bf 5.} Our analysis has revealed the strong likelihood that covariant 
quantization of QCD is inequivalent to canonical quantization, except perhaps 
with massless quarks. If both schemes lead to consistent theories, it means 
that QCD is not uniquely defined at the non-perturbative level.

In hindsight, the situation is not totally unexpected. We know of no way at 
present to show that the two quantization schemes are equivalent. As mentioned 
in the beginning, the QCD perturbation series does not uniquely define the 
theory. Feynman's one-loop equivalence was extended to all orders by DeWitt
\cite{DEWITT}. However, the proof relies on asymptotic fields (free quarks and 
gluons), so it is unlikely to extend to the non-perturbative regime. 
Mandelstam's proof \cite{MANDELSTAM} was given before the Gribov ambiguity was
discovered, so it does not take it into account. As well known, the Gribov 
ambiguity also spoils the Faddeev-Popov prescription except in some special 
cases \cite{FUJIKAWA}. Also, recent investigations \cite{SHABANOV} show that 
the application of the Batalin-Fradkin-Vilkovitsky theorem has its own 
shortcomings in the presence of the Gribov ambiguity.

In the last case in fact, BRST quantization has been shown to be inequivalent 
to Dirac quantization for certain models. The models, however, are quantum 
mechanical and do not have a topological term. In the following, we treat a gauge
theory with a topological term, where the inequivalence between covariant and 
canonical quantization can be explicitly seen.

\vskip .5cm
{\bf 6.} The theory is free electromagnetism in 1+1 dimensions, where `large'
gauge invariance is imposed in the canonical formalism, and a Fermi-type 
condition is used for covariant gauge. 
The analog of $\Xi$ is the electric field $F_{01}$.
For dimensional reasons, we normalize 
$\Xi=eF_{01}/2\pi$ so that ${\bf X}=(e/2\pi )\int dx^1 
A_1 (x)$, where $e$ is a non-zero parameter with the dimensions of charge 
(mass).

Gauge invariant quantization gives 
Maxwell equations $\partial_{\mu} F_{01} =0$ as operator relations, so
$F_{01}$ is a constant. This is compatible with the commutation relation
\be
[F_{\alpha\beta} (x) , F_{\mu\nu} (y) ] =i g_{\beta\nu} 
\partial_{\alpha}\partial_{\mu} 
D(x-y) \pm {\rm 3\,\, permutations}
\label{comm}
\ee
since the right hand side vanishes in 1+1 dimensions.

\vskip .25cm
{\it Canonical Quantization.}
In the $A_0=0$ gauge, the Lagrangian is 
\be
{\cal L}_{can} = \frac 12 (\partial_0 A^1)^2
- \frac {e\theta}{2\pi} \, \partial_0 A^1
\label{SW1}
\ee
The canonical conjugate of $A_1$ is $\Pi =-F_{01} -(e\theta )/2\pi$
and the Hamiltonian is
\be
H=\int dx^1 \frac 12 \left( \Pi + (\frac {e\theta}{2\pi})\right)^2
=\int dx^1 \frac 12 (F_{01})^2
\label{SW2}
\ee

The theory remains invariant under time-independent gauge transformations
\be
A_1 (t, x) \rightarrow A_1 (t, x) +\frac 1e  \lambda' (x)
\label{SW22}
\ee
generated by
\be
{\bf G} [\lambda ] = \frac 1e \int dx \, \lambda' (x) \,\Pi (t, x)
\label{SW3}
\ee
in order that the transformations do not change the physical state, 
$\Omega (x)=e^{i\lambda (x)}$ must approach one at infinity. The winding
number
\be
w [\Omega ] =&&-\frac i{2\pi} \int dx\, \Omega^{\dagger}(x) \Omega' (x) 
\nonumber\\=&&+
\frac 1{2\pi} (\lambda (+\infty )-\lambda (-\infty ) )
\label{SW4}
\ee
is then an integer and ${\bf X} =e\int dx A_1 (t, x) /2\pi$
transforms as ${\bf X}\rightarrow {\bf X} + w [\Omega ]$. 
We note that the theory has acquired another gauge-invariant dynamical
variable $e^{i2\pi {\bf X}}$ besides the electric field. 

The residual 
gauge-invariance (\ref{SW22}) implies that a constraint must be imposed.
Since the right hand side of (\ref{SW22}) is
$e^{-i{\bf G} [\lambda ]} A_1 (t, x) e^{i {\bf G} [\lambda ]}$, 
the desired constraint is
\be
e^{i{\bf G} [\lambda ]} |{\rm phys} > =|{\rm phys} >
\label{SW5}
\ee
where $\lambda$ is a gauge function with integer winding number, so that 
physical states are invariant under both `small' and `large' gauge transformations.
Since the electric field $F_{01}$, the Hamiltonian (\ref{SW2}), and the 
constraint $e^{i{\bf G} [\lambda ]}$ commute, it is easy to determine the 
physical spectrum. The eigenstate
$F_{01} (x) |\theta > = -(e\theta /2\pi )|\theta >$ is the physical ground 
state for $|\theta |<\pi$. For $\theta =\pm \pi$, there is 
a twofold degeneracy induced by $e^{\pm 2\pi i {\bf X}}$, and the structure 
repeats itself with a period $2\pi$.

On the other hand, if $\lambda$ is restricted to have zero winding number in 
(\ref{SW5}), $F_{01}=0$ is the ground state spectrum  for each
$\theta$. Evidently, `large' gauge invariance makes a difference. This 
concludes canonical quantization.

\vskip .25cm
{\it Covariant Quantization.}
In covariant gauge, the Lagrangian is
\be
{\cal L}_{cov} = \frac 12 (F_{01})^2 +\frac {e\theta}{2\pi} F_{01} 
 + \frac {\alpha}2 b^2 + b \partial_{\mu}A^{\mu} 
\label{SW6}
\ee
The canonical pairs are $(A^0, b)$ and $(A^1, \Pi) =(A^1, 
-F_{01}-(e\theta/2\pi))$ and the Hamiltonian is
\be
H=\int dx^1 \left( \frac 12
\left(\Pi + (\frac {e\theta}{2\pi})\right)^2
+A^0\partial_1\Pi -\frac {\alpha}2 b^2 -b\partial_1A^1\right)
\label{SW7}
\ee

One must also choose between the Fermi-type condition $b(x) |{\rm phys}>=0$
and the Gupta-Bleuler-type condition $b^{+} (x) |{\rm phys}>=0$. 
Both are acceptable for local observables since
$[b (x), F_{01} (y)]= [b^{+} (x), F_{01} (y)]= 0$. However, the Wightman 
function $<A_{\mu} (x) A_{\nu} (y)>_{\rm conn}$ is ill-defined  with the
Gupta-Bleuler condition, owing to infrared divergences peculiar to 1+1 
dimensions. Since the vacuum $|\theta >$ is normalizable, this means that 
$A_{\mu}$ is not well-defined as a quantum field. 

If we choose the Fermi-type condition, the equations of motion
\be
&&\partial_{\mu}F^{\mu\nu}=\partial^{\nu} b\nonumber\\
&&\Box b  =0
\label{SW8}
\ee
imply that the constraint is equivalent to
\be
&&b (0, x) |{\rm phys}>=0\nonumber\\
&&\partial_1\Pi (0, x) |{\rm phys} >=0
\label{SW9}
\ee
In this form it is clear that physical states are non-normalizable, so our
translational argument does not go through. However, the very definition of 
field operators as (tempered) distributions implies that the second equation in 
(\ref{SW9}) is equivalent to
\be
\int dx \, \Lambda' (x) \Pi (0, x) |{\rm phys}>=0
\label{SW10}
\ee
where the function $\Lambda (x)$ must fall off at infinity.
Hence `large' gauge invariance is not included, and the physical
ground state spectrum is $F_{01} =b=0$ for each $\theta$.

That $\theta$ does not affect the physics in general,  
may be demonstrated just as in 
section 3. The Hamiltonian (\ref{SW7}) for different vacuum angles is related 
by $H (\theta ) =e^{i\theta {\bf X}} H(0) e^{-i\theta {\bf X}}$. Also ${\bf X}$
commutes with the constraint (\ref{SW9}) at equal times 
$[{\bf X} (t) , b(t, x)]=[{\bf X}(t), \partial_1\Pi (t, x) ]=0$ hence all times.

In general, observables 
${\bf Q} (\theta )= e^{i\theta {\bf X}} {\bf Q}(0) e^{-i\theta{\bf X}}$
and physical states $e^{i\theta {\bf X}} |{\rm phys} 0>$ are in one-to-one 
correspondence, and the matrix elements are the same, so  physics is 
independent of $\theta$ in covariant gauge.  As before,
$H(\theta )=e^{i\theta {\bf X}} H(0) e^{-i\theta{\bf X}}$
in the canonical formalism also, but the proof does not go through there, since 
$e^{i\theta {\bf X}}$ does not commute with the `large' gauge constraint 
(\ref{SW5}) unless $\theta$ is an integer multiple of $2\pi$.

\vskip 0.25cm
{\it Summary.}
It is interesting that covariant and canonical quantization are inequivalent, 
even though there is no Gribov ambiguity (at least in Euclidean space). 
However, this may be deceptive. Path integrals for gauge theories remain 
undefined formal objects if only the fields and the action ${\bf I}$ are given. 
For QED in covariant gauge $\int 
[dA^{\mu}][db][d\psi][d\overline{\psi}]e^{i{\bf I}}$ does not distinguish 
whether Fermi or Gupta-Bleuler-type conditions are imposed. Similarly, if we 
take ${\bf I}=\int d^2x {\cal L}_{\rm can}$ of (\ref{SW5}), 
$\int [dA^1] e^{i{\bf I}}$ does not incorporate the constraint
(\ref{SW9}). If we put in boundary conditions or delta functionals to specify
the precise measure, we expect a Gribov-like obstacle in going from the 
canonical path integral to the covariant path integral. 

In any case, we see clearly how the constraints matters. Furthermore, 
lack of `large' gauge invariance does not imply lack of consistency.
We conclude that all the arguments point consistently in the direction 
that QCD is generally non-unique at the non-perturbative level. 

As for models, such as the the Schwinger model, the electroweak theory, or 
quantum gravity, we hope to discuss them in the future.

\vskip .5cm
{\bf 7.\,\,\,} 
If strong CP violation can be trivially solved in covariant gauge, it is 
natural to ask what our results have to say for the U(1) problem. Let us
briefly recall what the problem is and the various attempts to its solution. 
The axial U(1) symmetry of the QCD action is not apparent in nature, 
although a non-vanishing quark condensate suggests the existence of a ninth 
Nambu-Goldstone mode in the chiral limit \cite{GLASHOW}. This is known as the 
U(1) problem. The absence of such a mode in the hadronic spectrum is believed 
to be related to the chiral anomaly \cite{U1ANOMALY}. Based on the massless 
Schwinger model, Kogut and Susskind \cite{KOGUT} suggested that quark 
confinement solves the problem. Their idea was further taken up by Weinberg 
\cite{WEINBERG} and others \cite{KUGO}. With the discovery of instantons, 
't Hooft \cite{THOOFT} has suggested that they may provide a solution to the 
problem through the anomaly, but this has been challenged by Crewther 
\cite{CREWTHER}. Witten \cite{WITTEN} has proposed that the problem can be 
solved in the large $N_c$ (number of colors) limit, where the anomaly can be 
treated as a perturbation. Witten's proposal was later interpreted by Veneziano 
\cite{VENEZIANO} in terms of vector ghosts. Much of this discussion has been 
carried out using covariant gauge.

Fortunately, we only need a double Ward identity \cite{CREWTHER} to 
make our point. With $N_f$ flavors of quarks with equal mass $m$ for 
simplicity, the identity reads
\be
0 =&&-2i <\theta |\overline{q} q |\theta > +2i \frac{N^2_f}m \chi\nonumber\\
&&+ 2m \int d^4x <\theta | T^* \overline{q}i\gamma_5 q(x)
\overline{q} i\gamma_5 q (0) |\theta >_{conn}
\label{X6}
\ee
On the other hand, the absence of massless pseudoscalars for $m\neq 0$
in the flavor non-singlet channel gives
\be
0=&&\int d^4x\,\partial^{\mu}
<\theta | T^* \overline{q}\gamma_{\mu}\gamma_5\frac{\lambda^a}2 q(x)
\overline{q}i\gamma_5\lambda^b q(0) |\theta >\nonumber\\
=&&-\frac i2 <\theta | \overline q [\lambda^a,\lambda^b ]_+ q |\theta >
\nonumber\\
&&+m \int d^4x <\theta | T^* \overline{q}i\gamma_5 \lambda^a q(x) 
\overline{q} i\gamma_5 \lambda^b q (0) |\theta >_{conn} 
\label{X10}
\ee
If $\chi=0$ as in covariant gauge, (\ref{X6}) and (\ref{X10}) are essentially 
similar, and the same assumptions which imply the existence of Goldstone modes 
in the non-singlet channel as $m\rightarrow 0$ also imply the existence of a 
physical U(1) Goldstone mode as $m\rightarrow 0$.

One may object that the flavor singlet channel can mix with glueballs, whereas 
the non-singlet channel cannot, so the assumptions need not be the same. 
However, if $\chi=0$ and $<\theta | \overline{q} q |\theta >\neq 0$, the zero 
momentum propagator 
$\int d^4x <\theta | T^* \overline{q}i\gamma_5  q(x) 
\overline{q} i\gamma_5 (0) |\theta >_{conn}$ must become singular 
as $m\rightarrow 0$. Glueballs cannot generate such a singularity unless their 
masses also go to zero as $m\rightarrow 0$. Massless glueballs are as bad as
U(1) Goldstone modes.

To summarize, our results do not fundamentally alter the well-known trade-off 
between the U(1) problem and strong CP violation.
If anythingh, they emphasize it further.

\vglue 0.6cm
{\bf \noindent  Acknowledgements \hfil}
\vglue 0.4cm
This work was supported in part  by the US DOE grant DE-FG-88ER40388. 
We would like to thank many physicists in and out of Stony Brook and Tokyo
for discussions.

\vskip .5cm
{\bf Appendix}. In this short Appendix,
we would like to make some notes on n-vacua and 
normalizability. If we start out from n-vacua, the norm of the $\theta$ vacua 
is infinite as $<\theta |\theta >=\sum_n 1$. However, the n-vacua violate the 
superselection rule and cluster decomposition. Furthermore, the divergence
is merely the volume 
of the symmetry group, and may be factored out by a suitable choice of the 
configuration space. For example, a quantum mechanical system on the real line 
with a periodic potential $V(x+L)=V(x)$ is equivalent to a collection of 
systems on the interval $[0, L]$ with boundary conditions on the energy 
eigenfunctions $\psi (L) -e^{ikL} \psi (0) = \psi' (L) -e^{ikL} \psi' (0) =0$.
In the latter formulation, the ground state is normalizable for a given $k$.
(As a toy model for vacuum tunneling, $k$ corresponds to $\theta$.)

Similarly, the natural configuration space of pure Yang-Mills theory in the 
canonical formalism is ${\cal A}^3/{\cal G}^3$, where ${\cal A}^3$ is the space 
of (three) vector potentials $A_i^a (\vec x)$ and ${\cal G}^3$ is the group of 
proper gauge transformations $\Omega (\vec x )\rightarrow {\bf 1}$ with $|\vec 
x |\rightarrow \infty$. The classical vacuum must obey $F_{ij}^a (x) =0$. 
Again, all the n-vacua are identical in the classical limit, since ${\cal G}^3$ 
contains `large' as well as `small' gauge transformations. Instantons appear as 
noncontractible loops passing through the classical vacuum, corresponding to 
the fundamental group $\pi_1 ({\cal A}^3/{\cal G}^3) = Z$. Wave-functionals on 
${\cal A}^3/{\cal G}^3$ directly constitute the physical Hilbert space, so the 
vacuum functional has unit norm by axiomatics. However, $<\theta | K_{\mu} 
|\theta >$ remains ill-defined, since $K_{\mu}$ is not well-defined over
${\cal A}^3/{\cal G}^3$.

The conventional description based on n-vacua corresponds to taking the 
configuration space as ${\cal A}^3/{\cal G}_0^3$, where ${\cal G}_0^3$ is the 
group of `small' gauge transformations. Physical wave-functionals must then be 
invariant under `large' gauge transformations, so their norm is infinite.

\vskip 1cm
\setlength{\baselineskip}{15pt}

\end{document}